\documentclass{webofc}
\usepackage[varg]{txfonts}   
\usepackage{mathptmx}
\usepackage{amsmath}
\usepackage{amssymb}
\usepackage{amsfonts}
\usepackage{graphicx}
\usepackage{subfigure}
\newcommand{\myFigWidth}{0.48\textwidth}

\usepackage[linesnumbered,lined,boxed,commentsnumbered]{algorithm2e}

\begin{document}
\title{Global Thermodynamic Properties of Complex Spin\\ Systems Calculated from Density of States and \\ Indirectly by Thermodynamic Integration Method}
%
%

\author{\firstname{Marek} \lastname{Semjan}\inst{1}\fnsep\thanks{\email{marek.semjan@student.upjs.sk}} \and
\firstname{Milan} \lastname{\v{Z}ukovi\v{c}}\inst{1}\fnsep\thanks{\email{milan.zukovic@upjs.sk}} 
}

\institute{Institute of Physics, Faculty of Science, 
Pavol Jozef \v{S}af\'{a}rik University in Ko\v{s}ice,\\ \phantom{$^1$}Park Angelinum 9, 040 01 Ko\v{s}ice, Slovakia}

\abstract{%
Evaluation of global thermodynamic properties, such as the entropy or the free energy, of complex systems featuring a high degree of frustration or disorder is often desirable. Nevertheless, they cannot be measured directly in standard Monte Carlo simulation. Therefore, they are either evaluated indirectly from the directly measured quantities, for example by the thermodynamic integration method (TIM), or by applying more sophisticated simulation methods, such as the Wang-Landau (WL) algorithm, which can directly sample density of states. In the present investigation we compare the performance of the WL and TIM methods in terms of calculation of the entropy of an Ising antiferromagnetic system on a Kagome lattice -- a typical example of a complex spin system with high geometrical frustration resulting in a non-zero residual entropy the value of which is exactly known. It is found that in terms of accuracy the implementationally simpler TIM can deliver results comparable with the more involved WL method.
}
\maketitle
\section{Introduction}
\label{intro}
Calculation of global thermodynamic properties, which cannot be measured in Monte Carlo (MC) simulation directly, such as the free energy and the entropy, is generally a difficult task. A brute force approach of scanning the entire configuration space to obtain density of states is feasible only for sufficiently small systems. However, the exponential increase of the configurational space with the system size $N$ makes this approach intractable even for moderate sizes and small number of degrees of freedom, such as the Ising model with the configuration space increasing as $2^N$.

In statistical physics commonly used standard MC methods, such as the Metropolis algorithm (MA) \cite{metropolis1953}, allow direct evaluation of several thermodynamic quantities, such as the internal energy or magnetization, but not global ones, such as the free energy and the entropy. One possible approach that allows a direct evaluation of density of states (DOS) and consequently the entire thermodynamics is the Wang-Landau (WL) algorithm \cite{wang2001}, which has been successfully applied to a variety of problems, e.g., an efficient study of first-order and second-order phase transitions. It is a powerful tool for the investigation of systems with rough energy landscapes with large energy barriers separating local minima, which make the use of other standard methods infeasible. An alternative indirect approach to calculation of the global thermodynamic quantities, that avoids the calculation of DOS, is the so-called thermodynamic integration method (TIM) \cite{kirkpatrick1977}. Since its introduction in 1977 it has been sparingly used, even though several studies pointed to its competitiveness \cite{roma2004, zukovic2013}, for example in calculation of the ground-state entropy of some typical disordered/frustrated spin systems, such as the $\pm J$ Ising model and the spin-$s$ triangular lattice Ising antiferromagnet.

In the present study, we compare the performance of the WL and TIM methods in terms of calculation of global thermodynamic quantities of a highly frustrated Ising antiferromagnet on a Kagome lattice (IAKL)~\cite{syozi1951} with the focus on the entropy, the ground-state value of which is exactly known~\cite{kano1953}.

\section{Model and methods}
\label{sec-ma-wl}

\subsection{IAKL model}
\label{model}
The Hamiltonian of the studied spin $s=1/2$ IAKL system is given by
\begin{equation}
 \mathcal{H} = -J\sum_{\langle i,j\rangle }{\sigma_i\sigma_j}, \label{eq:H}
\end{equation}
where the first summation goes over the nearest neighbors and $\sigma_i = \pm 1$ is the spin at the $i$th site. In order to introduce frustration, interactions between neighboring spins were chosen to be antiferromagnetic ($J<0$). A schematic illustration of the Kagome lattice is shown in the inset of Fig.~\ref{fig:1a}. IAKL is a typical example of a complex spin system with high geometrical frustration resulting in a massive ground-state degeneracy with a finite residual entropy and no long-range ordering at any temperature. 

\subsection{MA and TIM}
\label{sec-ma}
MA is a well-known, general, easy to implement and therefore widely used MC method~\cite{metropolis1953}. The algorithm performs a random walk in the energy space. In every MC step a new state with the energy $\mathcal{H}$ is proposed and accepted with the probability $p(s_{\mathrm{old}} \rightarrow s_{\mathrm{new}}) = \min(1, e^{-\beta\Delta \mathcal{H}})$, where $\Delta \mathcal{H}$ is the energy difference between the new state and the old state, $\beta = 1/(k_BT)$ is the inverse temperature, and $k_B$ is the Boltzmann constant (hereafter set to $k_B=1$). MA can be used to directly calculate and investigate several quantities, such as the internal energy $e = \langle\mathcal{H} \rangle/N$ and magnetization $m = \langle M \rangle/N$, where $M = \sum_{i=1}^N{\sigma_i}$, $N$ is the number of spins and $\langle \dots \rangle$ denotes a thermal average. 

The entropy of the magnetic system with a discrete spin number $s$ can be obtained as a function of the inverse temperature by TIM \cite{kirkpatrick1977} as:
\begin{equation}
S(\beta) = N\ln{(2s+1)} + \beta E(\beta) - \int_0^\beta E(\beta')d\beta',\label{eq:ent} 
\end{equation}
where $E = N e$. Assuming equilibrium conditions, thermal averages calculated based on a fixed number of MC sweeps $N_{\mathrm{sweeps}}$, for a given temperature range $[T_{1},T_{N_T}]$ and a fixed lattice size $L$ the only relevant parameter that can influence accuracy of the entropy estimation is the temperature mesh density, characterized by the number of temperature points $N_T$ and their distribution. Generally, a denser mesh (large $N_T$) leads to a smaller quadrature error and thus a more accurate estimation.

\subsection{WL algorithm}
\label{sec-wl}
WL method is a relatively new MC method producing accurate results, including the global thermodynamic functions~\cite{wang2001}. A random walk is performed in the energy space to extract an estimate of DOS, $g(E)$, from which one can calculate the partition function at any temperature and consequently all other thermodynamic quantities. In particular, the partition function can be obtained as:
\begin{equation}
    Z(\beta) = \sum_{E}g(E)\exp(-\beta E),
\end{equation}
where the summation goes over all possible energy values $E$. Consequently, a mean value of any thermodynamic quantity, including the entropy, can be evaluated by using the standard statistical physics relations. For a given lattice size, the user-defined parameters in the WL method are the flatness criterion $F_C<1$ and the modification factor $f_{\mathrm{final}}>1$~\cite{wang2001}. Generally, the closer are the values of $F_C$ and $f_{\mathrm{final}}$ to 1.0 the more precise results can be expected. Typically chosen values include $F_C=0.8$ or $0.9$ and $f_{\mathrm{final}}=1+10^{-k}$, for $k=8$, 9, and 10.

\begin{figure}[t!]
\centering
\subfigure{\label{fig:1a}\includegraphics[width=\myFigWidth,clip]{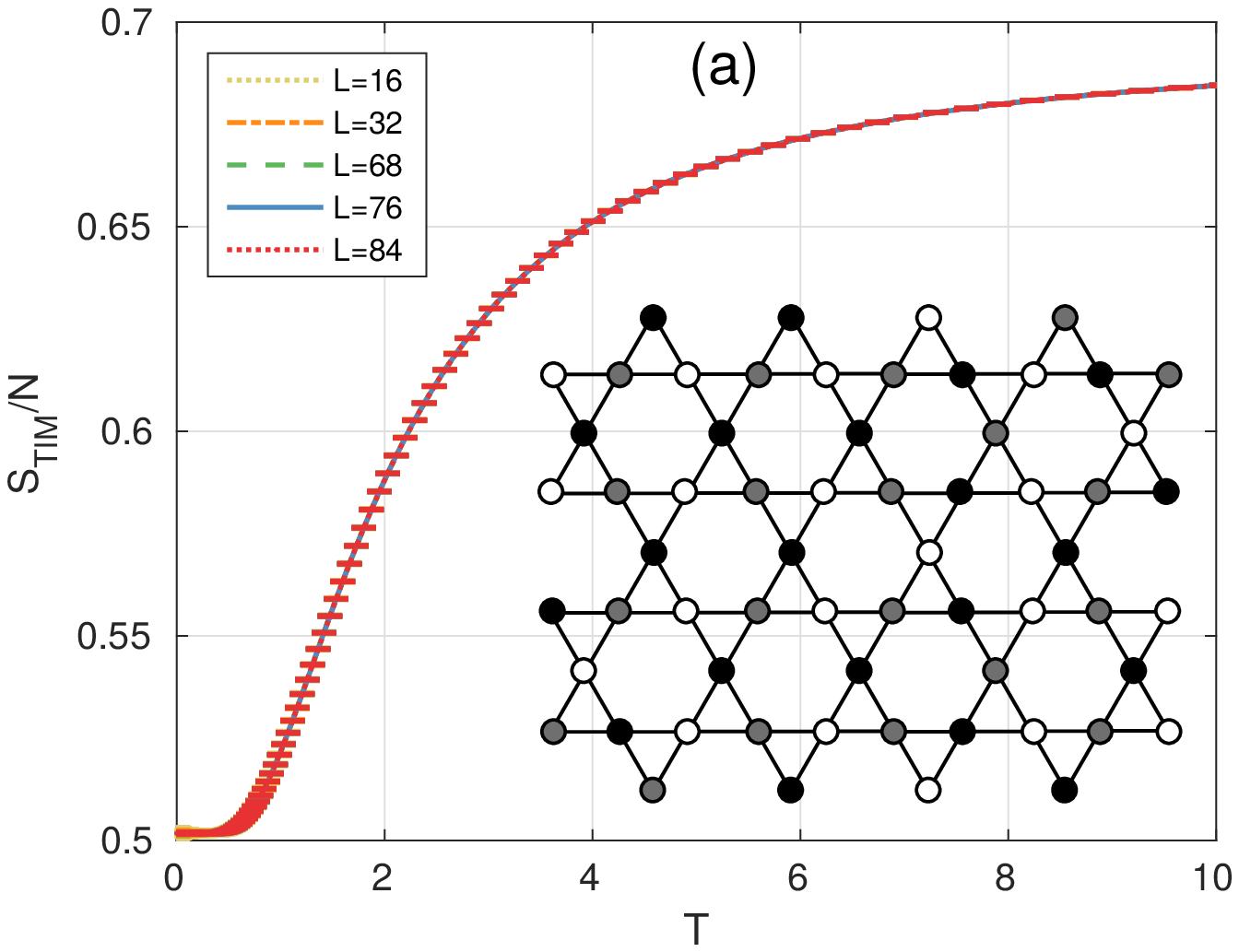}}
\subfigure{\label{fig:1b}\includegraphics[width=\myFigWidth,clip]{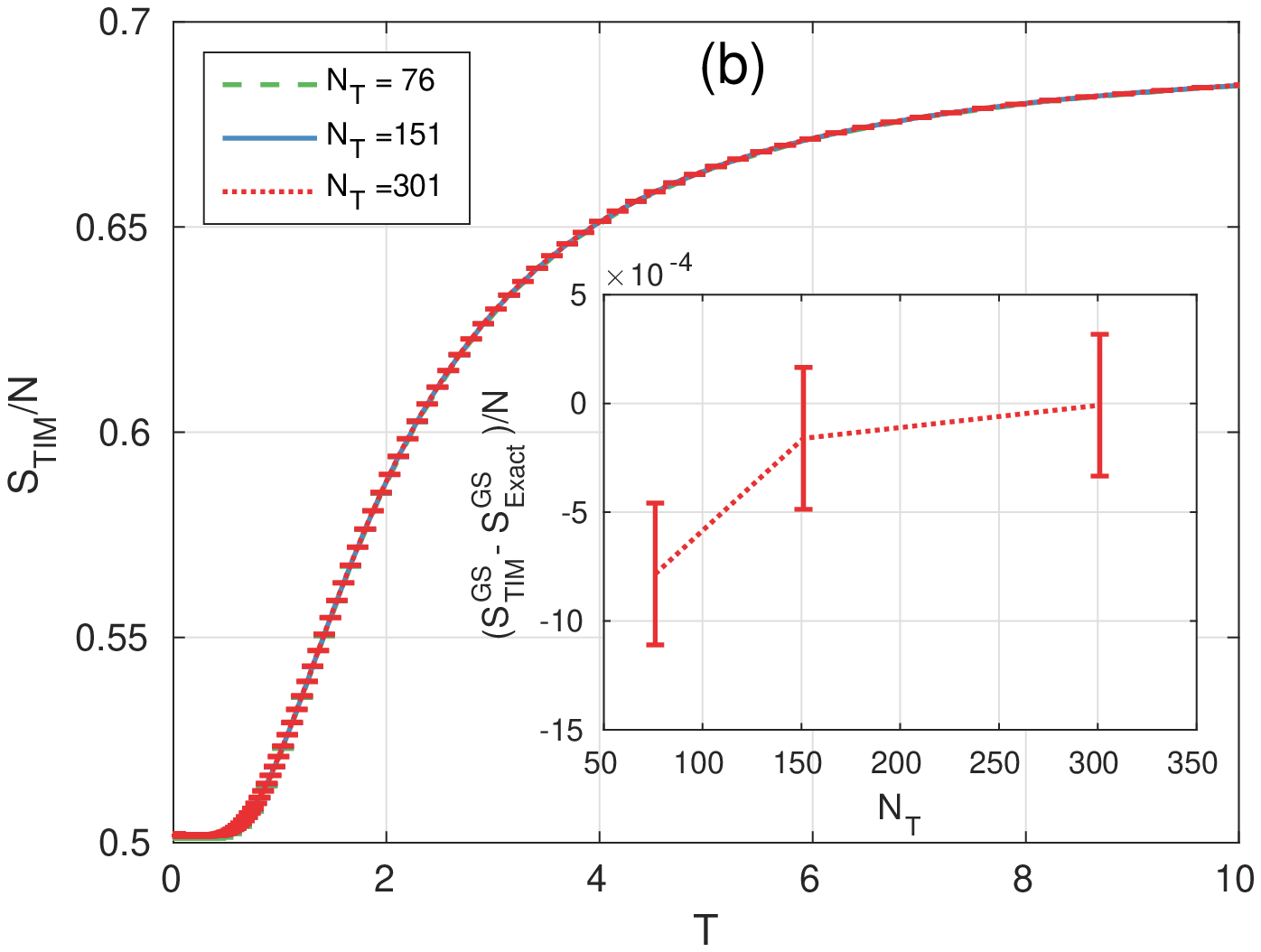}}
\caption{(a) Entropy density obtained by the TIM method, for different $L$ with $N_T=301$. The inset shows the Kagome lattice composed of three interpenetrating sublattices (different shades). (b) Entropy density for $L=32$ and different values of $N_T$. The inset shows difference between the ground state entropy density calculated using TIM and the exact value.}
\label{fig:Ldep}       
\end{figure}

\section{Results and discussion}
\label{results}
We performed several simulations using both MA and WL to calculate the entropy per spin (entropy density) and the free energy. Below, we only present the former quantity, as the latter one is just a simple function of the former. The presented results were calculated as averages obtained from 10 independent runs. 

Fist, we present results obtained by MA, with the following parameters: $N_{\mathrm{sweeps}} = 5\times10^5$ and $[T_{1},T_{N_T}]=[0.0064,\infty]$ (or $[\beta_{1},\beta_{N_T}]=[156.3731,0]$). The temperature mesh was chosen non-uniformly with the largest density of points in the region of the largest variation of the energy $E(\beta)$ in order to increase the precision of the numerical quadrature (Eq.~(\ref{eq:ent})). In Figure~\ref{fig:1a} we examine the lattice size dependence of TIM with $N_T = 301$, for $L=16,\ldots,84$. All the curves are found to collapse on a single curve within the error bars and, hence, we can conclude that the finite-size effects are very small. In Figure~\ref{fig:1b} we study the influence of the parameter $N_T$, for a fixed value of $L=32$. $N_T=151$ and 76 case were obtained from the initial $N_T=301$ cases by repeatedly removing every second nod from the previous denser grid. Again, all the curve appear to coincide within statistical errors. Nevertheless, as shown in the inset, by comparing the ground-state value estimate $S^{GS}_{MA}/N$ with the exact value $S^{GS}_{\mathrm{exact}}/N=0.5018$~\cite{kano1953} one can notice a gradual improvement with the increasing $N_T$, even though accuracy is fairly high for all the values of $N_T$.

\begin{figure}[t!]
\centering
\subfigure{\label{fig:2a}\includegraphics[width=\myFigWidth,clip]{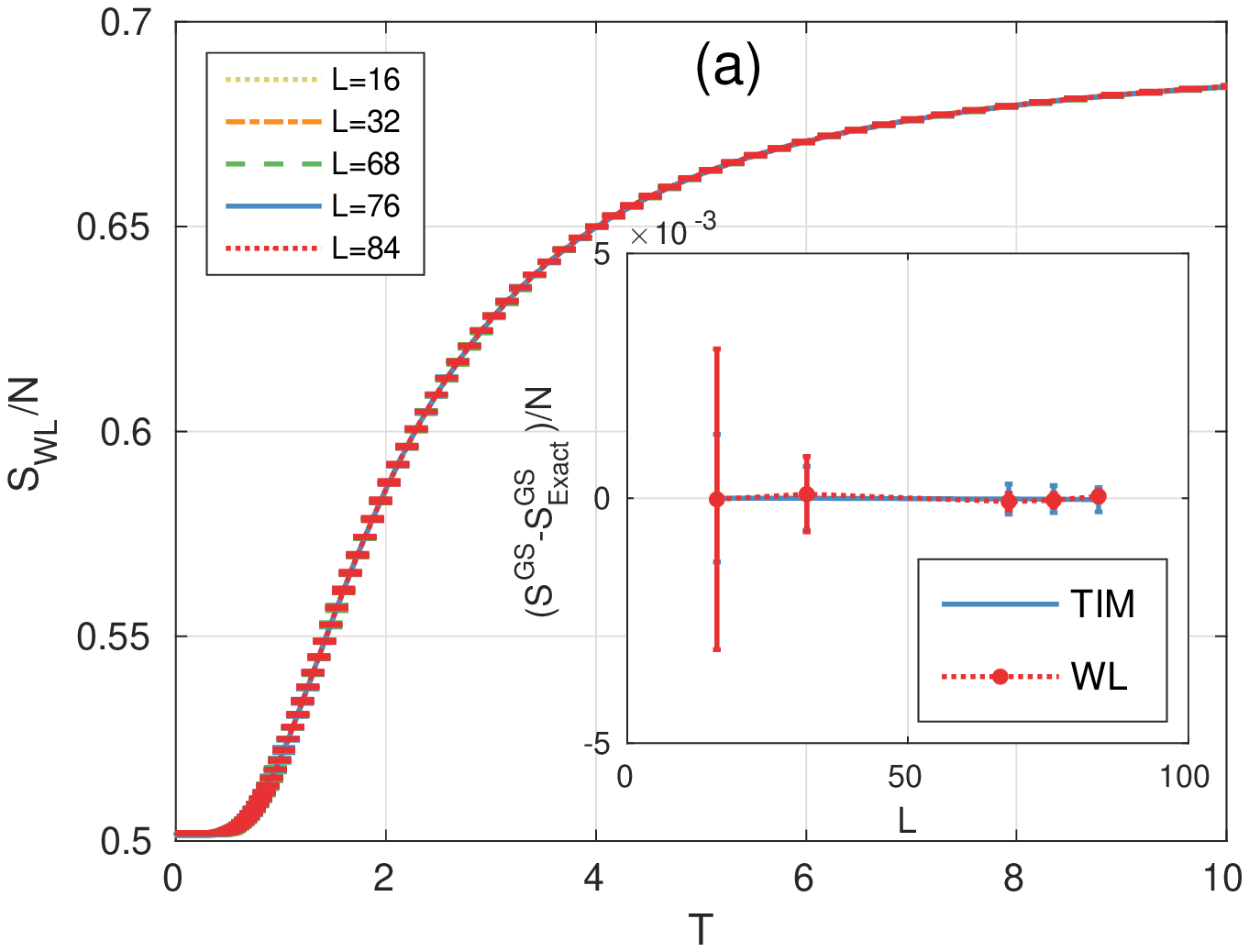}}
\subfigure{\label{fig:2b}\includegraphics[width=\myFigWidth,clip]{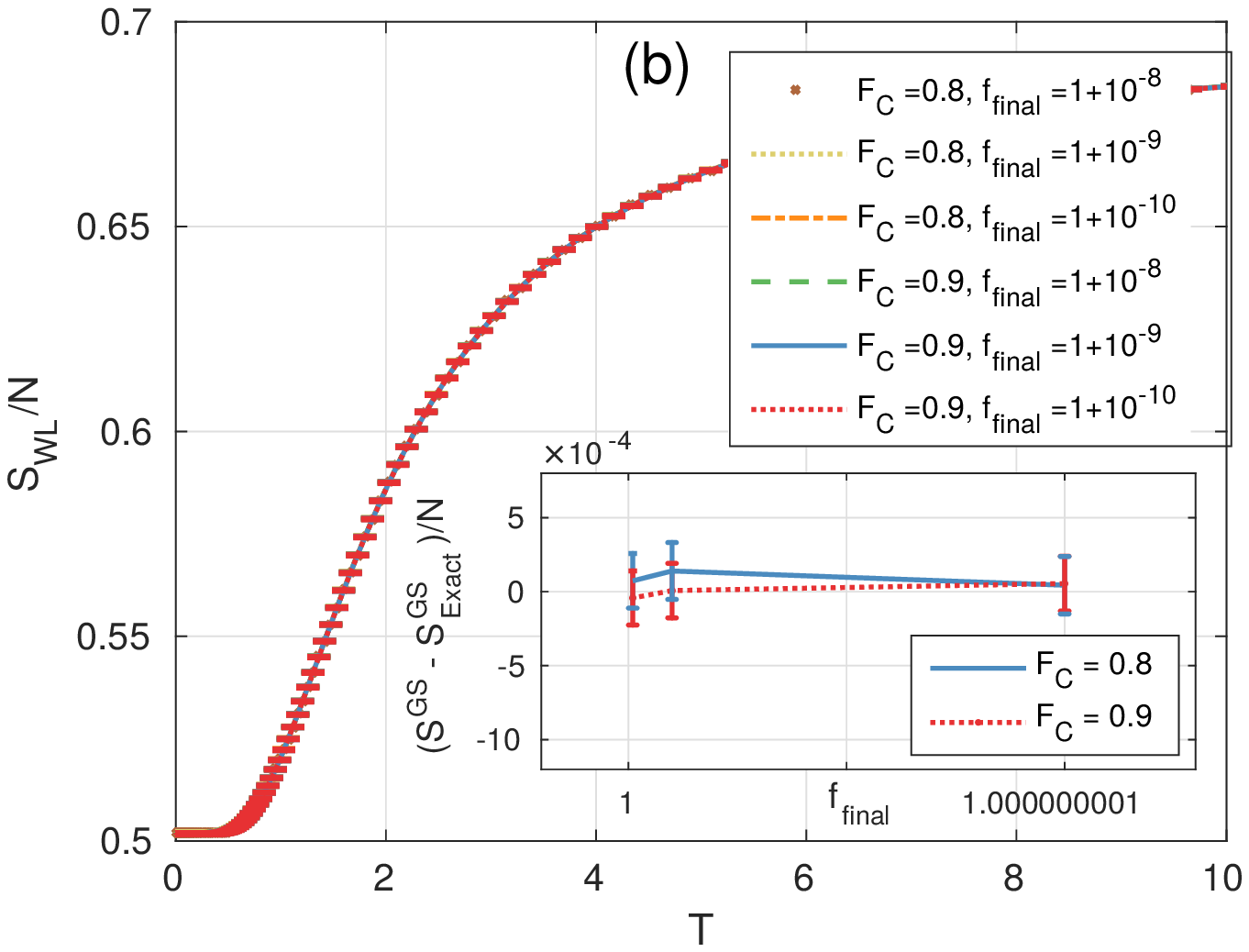}}
\caption{(a) Entropy density obtained by the WL method, for different $L$ with $F_C = 0.8$ and $f_{\mathrm{final}} = 1 + 10^{-8}$. The inset shows differences of the residual values by the respective methods from the exact value. (b) Entropy density curves for different values of the WL parameters $F_C$ and $f_{\mathrm{final}}$ and their difference from the exact value in the ground state (inset).}
\label{fig:wlDep}       
\end{figure}

In Figure~\ref{fig:2a} we present the WL results for different $L=16,\ldots,84$ and the simulation parameters set to $F_C = 0.8$ and $f_{\mathrm{final}} = 1 + 10^{-8}$. Like for the TIM results in Figure~\ref{fig:1a}, all the curves coincide within the error bars. There is also a good coincidence between the WL and TIM results, as evidenced in the inset of Figure~\ref{fig:2a} that shows the deviation of the residual entropies obtained by the WL method and TIM (for $N_T=301$) from the exact value. Finally, in Figure~\ref{fig:2b} we demonstrate the effect of the choice of the WL method parameters $F_C$ and $f_{\mathrm{final}}$. Again, for the standard values the differences are within the error bars. 

To conclude, when the standard values of parameters are chosen, both TIM and WL methods deliver sufficiently accurate results that mutually indistinguishable. The advantage of the TIM is its implementational simplicity but the simulation has to be run at all desired temperatures. On the other hand, having obtained DOS from the WL simulation enables straightforward calculation of any thermodynamic quantity at all temperatures.

\subsection*{Acknowledgement}
This work was supported by the Scientific Grant Agency VEGA (Grant No. 1/0531/19).

%
%
%

\end{document}